\documentclass[12pt]{article}

\usepackage[margin=3cm]{geometry}

\usepackage{amsmath}
\usepackage{amssymb}
\usepackage{algorithm}
\usepackage{algpseudocode}
\usepackage{url}
\usepackage{setspace}
\usepackage{graphicx}
\usepackage{wasysym}			
\usepackage{pdfpages}

\newcommand{\R}{\mathbb{R}}
\newcommand{\N}{\mathbb{N}}

\newcommand{\dd}{\partial}
\newcommand{\Om}{\Omega}

\newcommand{\lmb}{{\lambda}}
\newcommand{\sig}{\sigma}
\newcommand{\sigb}{\bar{\sigma}}
\newcommand{\kap}{\kappa}

\begin{document}

%
%

%
%
\title{The Idea and Concept of\\ {\sc Metos3D} -- A Marine Ecosystem Toolkit for Optimization and Simulation in 3-D}
\author{Jaroslaw Piwonski \thanks{\texttt{jpi@informatik.uni-kiel.de}}, Thomas Slawig \thanks{\texttt{ts@informatik.uni-kiel.de},
both: Department of Computer Science, Algorithmic Optimal Control -- $CO_2$ Uptake of the Ocean,
Excellence Cluster The Future Ocean, Christian-Albrechts-Platz 4, 24118 Kiel, Germany.}}
\date{Reissue of first version 2010}

\maketitle

%
%
\begin{abstract}
The simulation and parameter optimization of coupled ocean circulation and ecosystem models in
three space dimensions is one of the most challenging tasks in numerical climate research.
Here we present a scientific toolkit that aims at supporting researchers by defining clear coupling interfaces,
providing state-of-the-art numerical methods for simulation, parallelization and optimization while using only
freely available and (to a great extend) platform-independent software.
Besides defining a user-friendly coupling interface (API)  for marine ecosystem or biogeochemical models,
we heavily rely on the Portable, Extensible Toolkit for Scientific computation (PETSc) developed in
Argonne Nat. Lab. \cite{PETSc} for a wide variety of parallel linear and non-linear solvers and optimizers.
We specifically focus on the usage of matrix-free Newton-Krylov methods for the fast computation of
steady periodic solutions, and make use of the Transport Matrix Method (TMM) 
introduced by Khatiwala et al. in \cite{KhViCa05}.
\end{abstract}

%
%

%
%
\section{Introduction}

Here we describe the main features of coupled ocean circulation and  ecosystem models and
how {\sc Metos3D} will help to flexibly perform  simulation and optimization runs
for a wide range of model configurations.


\subsection{Coupled Ocean Circulation and Ecosystem Models and Simulations}

Marine ecosystem models play an important  role for the investigation of the global  carbon cycle. 
The ocean water takes up a huge amount of $CO_2$ from the atmosphere,
and the question how long  the carbon (or carbon dioxide) remains in the ocean 
-- especially in the deeper layers -- clearly 
effects the future uptake and thus also prognoses on climate change.
Coupled ocean circulation and  ecosystem models 
consist of two parts, namely the ocean circulation itself and
the biogeochemistry describing (or trying to describe) the interplay between
marine chemistry, marine biology and the geological conditions and processes.
The main coupling direction between  both parts (at least the one that in many cases is considered only) is the one from
ocean circulation to biogeochemistry, whereas the inverse coupling is less clear and
often ignored in coupled models and simulations. When running biogeochemistry model with
given ocean circulation without taking into account this inverse effect, sometimes the notions
{\em offline computation}\/ or (for the simulated quantities) {\em passive tracers}\/ is used.

Whereas for the ocean circulation the governing equations are quite clear
(Navier-Stokes equations, temperature/energy equation, and transport equation for salinity),
for biogeochemical models there is a big variety. The models can have from two up to dozen of
tracer variables, and consequently a similar different number of transport equations.
Moreover, many model parameters (as growth or dying rates) are not directly measurable.
Therefore there is a high demand on parameter optimization to obtain those that fit the model
output to given observational data, and furthermore
a need for a possibility to quickly examine
changes within parameters, parameterizations or even
number of tracers and model structure.
 Preferably this should be done
within a three dimensional physics, on a global scale
and using circulation from an up-to-date ocean circulation or global climate model.

When performing a coupled simulation,  both parts (i.e. the simulation of the ocean circulation and
of a biogeochemical model) usually  require a huge amount of  computing time, 
especially  in three space dimensions. In a typical case,
a coupled run from a spatially constant tracer distribution to a steady periodic solution
takes thousands of years of model time. 


\subsection{Idea and Structure of {\sc Metos3D}}

The idea of {\sc Metos3D} is to provide a toolkit that has the following properties: 
\begin{itemize}
\item It allows a flexible coupling of the ocean model component and the biogeochemistry part for simulation. 
	In the first place it provides a coupling to a transport matrix method introduced by Khatiwala et al., see \cite{KhViCa05}.
\item It provides an interface/API between a quite general class of marine biogeochemical models that
	can be coupled to the ocean model part and to the simulation and optimization methods.  
	It offers the opportunity to use tools of Algorithmic/Automatic Differentiation (AD, see e.g. \cite{GriWal08}) for
	simulation, sensitivity computations and optimization.
\item It provides a Newton-based method to compute steady periodic solutions as well as
	a classical pseudo time-stepping spin-up. 
	It deploys state-of-the-art techniques of scientific computing including equation solvers, optimization routines and
	parallelization techniques to accelerate all simulation and optimization runs.
	Here {\sc Metos3D} is based on the well-established {\sc PETSc} library, see \cite{PETSc}.
\item It relies only on software which is freely and public available and runs on different platforms. 
\item The implementation is platform-independent and allows coupling of models written in C and {\sc Fortran}.
\end{itemize}


%
%
\section{Marine Ecosystem Tracer Simulation}

Marine biogeochemical tracers are substances in the ocean water that are undergoing chemical or biochemical reactions.
The processes in this ecosystem are thus governed by time dependent transport equations with additional, usual non-linear coupling terms.
Since the ocean circulation determines the tracer transport, solving these equations means
solving the Navier-Stokes equations to obtain the turbulent mixing/diffusion  coefficients  and the velocity field
for advection.



\subsection{Transport Equations}

The system of  transport equations for the vector of  tracers $y=(y_i)_{i=1,\ldots,n}$ generally reads
\begin{align}
\label{DiffAdEquation}
\frac{\partial y_i}{\partial t} & = \nabla \cdot (\kap \nabla y_i) - \nabla y_i \cdot v + q(y,u),
\quad i=1,\ldots,n,
\end{align}
where  $v$ is the  velocity vector and $\kap$ the turbulent mixing/diffusion coefficient.
Both have to satisfy the Navier-Stokes equations which can be solved simultaneously
with (\ref{DiffAdEquation}) or  (in the so-called {\em offline mode}) in advance.
 
The usual non-linear coupling term $q$ (also called {\em source minus sink}\/ or {\em SMS term})
models the interaction between the tracers, e.g. photosynthesis and growth and dying or
consumption processes. Their modeling involve several only heuristically determined parameters,
here summarized in a vector $u$, that are often the subject of parameter optimization in order to fit the model to given data.

\subsection{Marine Ecosystem Models}

There is a wide range of marine  biogeochemical  or ecosystem models. They vary in the number of tracers and
thus equations, and moreover in the number and types of processes they model and the way {\em how}\/ they do this.
In these {\em parametrizations}\/ the model parameter vector $u$ plays an important role:
The number and meaning of the parameters differ from model to model.
To handle a wide range of ecosystem models thus requires a high flexibility in the number of both tracers and parameters.

Examples for tracers are nitrate, phosphate, chlorofluorocarbons
or carbon dioxide. Modeling the carbon cycle for example requires describing
alkalinity, a complex reaction of inorganic tracers.
A detailed description of this reaction can be found in the online document
about the design of OCMIP-2 simulations
(Ocean Carbon-Cycle Model Intercomparison Project, http://www.ipsl.jussieu.fr/OCMIP/phase2/simulations/design.ps).

Phosphate on the other hand is a major nutrient for ocean biota (organisms) which is
represented by organic tracers in biogeochemical models, either implicitly by
dissolved organic phosphorus, nitrogen or carbon or explicitly by 
phytoplankton or zooplankton. Thus in this case the intent of the model equations is
to describe processes like nutrient consumption, growing, grazing or dying
rather than chemical reactions. 

%

Most models also admit terms representing sinking of inorganic tracers, like iron or calcite
particles, and organic tracers, like detritus (dead matter) or particulate organic phosphorus.
Moreover all these processes depend on many external geographic conditions, i.e.
precipitation, insolation, air pressure or latitude.

In the following we will describe two simple global models based on inorganic phosphate
($PO^{-3}_4$, or $PO4$ for short) as an example.


\subsubsection{Example 1: The PO4-DOP Model}
\label{PO4DOP}

The PO4-DOP model admits phosphate and dissolved organic phosphorus (DOP).
The tracers are denoted by $ y = (y_1, y_2)^T = (y_{PO4}, y_{DOP})^T$.
The biological production (the net community productivity) is calculated as a function
$f_1$ of nutrients and light $I$. The production is limited using a 
half saturation function, also known as Michaelis-Menten kinetics,
and a maximum production rate parameter $\alpha$.
\begin{align*}
f_1(y_1,I)	& = \alpha \, \frac{y_1}{y_1 + K_1} \, \frac{I}{I+K_I}
\end{align*}
Light, here, is a portion of short wave radiation $I_{SWR}$, which is computed as a
function of latitude and season following the astronomical formula of Paltridge and Platt \cite{PalPla76}.
The portion depends on the photo-synthetically available radiation $\sig_{PAR}$,
the ice cover $\sig_{ice}$ and the exponential attenuation of water.
\begin{align*}
I		& = I_{SWR} \, \sig_{PAR} \, (1 - \sig_{ice}) \, \exp( - z \, K_{H2O} )
\end{align*}
A fraction of the biological production $\sig_2$ remains suspended in the water column
as dissolved organic phosphorus, which remineralizes with a rate $\lmb'_2$.
The remainder of the production sinks as particulate to depth where it is
remineralized according to the empirical power law relationship determined by
Martin et al. \cite{MaKnKaBr87}. Similar descriptions for biological production can be
found in \cite{PaFoBo05}, \cite{DuFoPa05} and \cite{YamTaj97}.

Moreover the model formulation consists of a production (sun lit, euphotic) zone,
with a depth of $l'$, and a
noneuphotic zone, $\Om_1$ and $\Om_2$ respectively.
The equations read
\begin{align*}
q_1(y) & = - f_1(y_1,I)			 + \lmb'_2 \, y_2 & \text{in $\Om_1$} \\
q_1(y) & = + \sigb_2 \, \dd_z \, F_1(y_1,I) + \lmb'_2 \, y_2 & \text{in $\Om_2$} \\
 & \\
q_2(y) & = + \sig_2 \, f_1(y_1,I) 	 - \lmb'_2 \, y_2 & \text{in $\Om_1$} \\
q_2(y) & = 				 - \lmb'_2 \, y_2 & \text{in $\Om_2$,}
\end{align*}
where
\begin{align*}
F_1(y_1,I)	& = (z/l')^{-b} \int_0^{l'} \! f_1(y_1,I) \, d\xi \,.
\end{align*}
The following table summarizes the parameters within the PO4-DOP model:
\begin{center}
\begin{tabular}{c|l|c}
Symbol		& Description								& Unit \\ \hline
$\alpha$		& maximum community production rate			& $ 1 / d $ \\
$K_{H2O}$	& attenuation of water						& $ 1 / m $ \\
$K_1$		& half saturation constant of PO4 				& $ m \, mol P / m^3 $ \\
$K_I$		& half saturation constant of light	 			& $ W / m^2 $ \\
$\lmb'_2$		& remineralization rate of DOP 					& $1 / d$ \\
$\sig_2$ 		& fraction of DOP, $\sigb_2 = (1-\sig_2)$ 			& $ - $ \\
$b$			& sinking velocity exponent					& $ - $
\end{tabular}
\end{center}


\subsubsection{Example 2: The PO4-DOP-PHY Model}

The PO4-DOP-PHY model admits explicitly modeled phytoplankton (PHY),
which is implicitly prescribed within the $\alpha$ parameter in the former model
for example. The tracers are denoted by $y = (y_1, y_2, y_3)^T = (y_{PO4}, y_{DOP}, y_{PHY})^T$.

The biological production additionally depends now on the phytoplankton concentration and
a growth rate $\mu_3$
\begin{align*}
f_1(y_1,y_3,I) & = \mu_1 \, y_3 \, \frac{y_1}{y_1 + K_1} \, \frac{I}{I+K_I} ,
\end{align*}
while the computation of light do not change.
The fraction of phytoplankton that is grazed is controlled by $\sig_3$.
Additionally there are three loss rates $\lmb_3, \lmb'_3$ and $\kap_3$ for phytoplankton. 
They can be considered as parameterization of various processes like mortality rate,
aggregation of cells and immediate sinking or viral infections.
Their detailed description can be found in Kriest et al \cite{KrKhOs10}.
The model equations read
\begin{align*}
q_1(y) & = - f_1(y_1, y_3,I) + \lmb'_2 \, y_2 & \Om_1 \\
q_1(y) & = + \sigb_2 \, \dd_z \, F_3(y_1,y_3,I) + \lmb'_2 \, y_2 & \Om_2 \\
 & \\
q_2(y) & = + \sig_2 \, f_3(y_1,y_3,I) + \lmb_3 \, y_3 + \kap_3 \, y_3^2 - \lmb'_2 \, y_2 + \lmb'_3 \, y_3 & \Om_1 \\
q_2(y) & = - \lmb'_2 \, y_2 + \lmb'_3 \, y_3 & \Om_2 \\
 & \\
q_3(y) & = \sigb_3 \, f_1(y_1,y_3,I) - \lmb_3 \, y_3 - \kap_3 \, y_3^2 - \lmb'_3 \, y_3 & \Om_1 \\
q_3(y) & = - \lmb'_3 \, y_3 & \Om_2 \,,
\end{align*}
where
\begin{align*}
f_3(y_1,y_3,I) & = \sig_3 \, f_1(y_1, y_3,I) \\
F_3(y_1,y_3,I) & = (z/l')^{-b} \int_0^{l'} \! f_3(y_1,y_3,I) \, d\xi \,.
\end{align*}
Again, the following table summarizes the model parameters:
\begin{center}
\begin{tabular}{c|l|c}
Symbol		& Description							& Unit \\ \hline
$\mu_1$		& maximum growth rate of PO4				& $ 1 / d $ \\
$K_{H2O}$	& attenuation of water					& $ 1 / m $ \\
$K_1$		& half saturation constant of PO4 			& $ m \, mol P / m^3 $ \\
$K_I$		& half saturation constant of insolation		& $ W / m^2 $ \\
$\lmb'_2$		& remineralization rate of DOP 				& $ 1 / d $ \\
$\sig_2$ 		& fraction of DOP, $\sigb_2 = (1-\sig_2)$ 		& $ - $ \\
$\lmb_3$		& loss rate of PHY, euphotic zone			& $ 1 / d $ \\
$\kap_3$		& quadratic loss rate of PHY, euphotic zone	& $ 1 / d \cdot m^3 / m \, mol P $ \\
$\lmb'_3$		& loss rate of PHY, whole water column		& $ 1 / d $ \\
$\sig_3$ 		& grazing fraction, $\sigb_3 = (1-\sig_3)$ 		& $ - $ \\
$b$			& sinking velocity exponent				& $ - $ 
\end{tabular}
\end{center}


%
%
\section{Transport Matrix Approach}

In the first place we provide the pre-computed ocean circulation fields by so-called {\em transport matrices}.
In \cite{KhViCa05} Khatiwala et al. presented an approach well suited for
passive tracer simulation. Their method is a good trade-off between
complexity and accuracy. Moreover the assumption they make, i.e.
important biogeochemical processes happen within a water column
(and a small zone around it),
matches ours for the biogeochemical interface (see Section \ref{BGCInterfaceAssumption}).

The idea is based on the linearity of the time dependent diffusion and advection operators.
The system of transport equations can always be written as
\begin{align*}
\frac{\partial}{\partial t} y(t) & = A(t) \, y(t) + q(t,y(t),u) \,
\end{align*}
where $A$ admits both, diffusion and advection, and becomes a block diagonal matrix after
discretization. Due to the locality of the differential operators the matrices remain extremely
sparse for large systems of equations. Using basis functions, $A$ can be evaluated and stored.

In doing so several aspects have to be taken into account.
The choice of appropriate basis functions is crucial to avoid too sharp gradients or too high diffusivity.
Moreover modern ocean circulation
models consist of operator splitting schemes to tackle the different spacial extents of
the ocean. Operators effecting the longitudinal and latitudinal directions
are treated explicitly, while vertical effects are computed implicitly in time. This results
in two different types of matrices.

A compromise has to be found for storage too. Here Khatiwala et al. propose
monthly averaged diffusion and advection matrices, which can be interpolated
during computation.
Their experiments show that twelve explicit and twelve implicit
matrices provide a sufficient accuracy at minimal storage requirements.

Nevertheless, using a transport matrix approach, the whole coupled simulation results in a scheme of
matrix-vector multiplications and evaluations of the biogeochemical model. Hence
\begin{align*}
y_{i+1} & = A_{imp,i} \, (A_{exp,i} \, y_i + q_i(y_i,u)), \quad i = 0, \dots, n_{step} - 1 \, .
\end{align*}
Its simplicity is clearly an advantage compared to other couplings. There is still a resolved
seasonal cycle and a sparse matrix-vector multiplication is a well supported standard
operation in numerical libraries. Moreover no diffusion and advection have to be implemented.


%
%
\section{Spin Up and Computation of Periodic Solutions}
\label{sec:spinup}
Transport equations  -- after spatial discretization -- can be viewed as a subclass of systems of ordinary differential equations. They are in most cases solved by a time-integration procedure, i.e. an Euler or Runge-Kutta type method.
Given a coupling term $q$ the solution depends on the so called \emph{forcing}, i.e.
the turbulent mixing/diffusion coefficient, the velocity vector, wind, insolation etc., which  can be time dependent. These data are either given or can be (pre-)computed by an ocean circulation model. In biogeosciences it is common
to search for solutions forced by climatological (time dependent but periodic) data.
Here the notion \emph{dynamical spin up problem} or (for the solution)
\emph{equilibrium} is used.

For a given time interval $[0,T]$ the process of integration over the whole interval
from an initial value $y(0)$ defines a mapping $\phi$ with
\begin{align*}
y(T) & = \phi( y(0) ) \, ,
\end{align*}
while for the solution the periodic property reads
\begin{align*}
y(t+T) & = y(t), \quad t \in [0,T] \, .
\end{align*}
Thus a periodic solution $y$ is a fixed point for the -- in general nonlinear -- mapping $\phi$.

%
%
%
%
%


%
%
\section{Biogeochemical Model Interface}

As indicated in the former sections marine ecosystem simulations are complex
at all points. On the one hand they are computationally costly, on the other hand
they admit a large variety of biogeochemical models.
Especially the latter highlights a need for the possibility of quickly examine
new model formulations and parameterizations.
Consequently our intent is to propose an interface for coupling ocean physics to
biogeochemical models. 

\subsection{Assumptions and Description}
\label{BGCInterfaceAssumption}

The main assumption we make is that, regarding biogeochemistry, the most important processes
happen within a water column, even in three space dimensions. This assumption is not definitive and
can change in the future, but seems to be reasonable in the beginning. It simplifies the
implementation of a model a lot and, as we will see, does not constrain the interface too much.
Thus the interface admits, besides the number of layers and tracers, $n_{layer}$ and $n_{tracer}$
respectively, an array of given tracer concentrations for a water column
$ ((y_{j,i})_{i=1}^{n_{layer}})_{j=1}^{n_{tracer}}$ and an array for computed tracer concentrations
by the model $ ((y_{bgc,j,i})_{i=1}^{n_{layer}})_{j=1}^{n_{tracer}} $.

The second assumption made is that the year consists of 360 days. This becomes of importance when
regarding the ocean time step $\Delta t$ and the point in time $t$. Moreover it has an effect on the
provided, precomputed short wave radiation $I_{SW\!R}$.

Considering light computations from the latter,
see Section \ref{PO4DOP}, the interface additionally provides the fraction of day $\sig_{day}$,
the ice cover $\sig_{ice}$ as well as the heights $ (h_i)_{i=1}^{n_{layer}} $ and
depths $ (d_i)_{i=1}^{n_{layer}} $ for the current water column.

Another important aspect in biogeochemical models is the air-sea gas transfer. It depends on
the mentioned ice cover and the wind speed $ v_{wind} $ on the surface. Moreover
the latitude $ \theta $ of the water column, the water temperature $ (T_i)_{i=1}^{n_{layer}} $
and salinity $ (S_i)_{i=1}^{n_{layer}} $ are needed for its computation.

Last by not least the interface provides a possibility to pass parameters to the model.
Their number is denoted by $ n_{param} $ and the array by $ (u_{i})_{i=1}^{n_{param}} $.

The following table summarizes the input and output arguments of the interface,
their notations, ranges, units and short descriptions.

\begin{center}
\begin{tabular}{l|l|c|l}
Notation										& Range				& Unit		& Description \\ \hline
$ t $											& $ \in [0,1] $			& -			& point in time \\
$ \Delta t $									& $ \in [0,1] $			& -			& ocean time step \\
$ \theta $										& $ \in [-90, 90] $		& $ {}^\circ $	& latitude \\
$ I_{SWR} $									& $ \in [0, \infty) $		& $ W/m^2 $	& short wave radiation \\
$ \sig_{day} $									& $ \in [0,1] $			& -			& fraction of day \\
$ \sig_{ice} $									& $ \in [0,1] $			& -			& ice cover \\
$ v_{wind} $									& $ \in [0, \infty) $		& $ m/s $		& wind speed \\
$ n_{tracer} $									& $ \in \N $			& -			& number of tracers \\
$ n_{layer} $									& $ \in \N $			& -			& number of layers  \\
$ (h_i)_{i=1}^{n_{layer}} $							& $ \in \R^{n_{layer}} $	& $ m $		& array of layer heights \\
$ (d_i)_{i=1}^{n_{layer}} $							& $ \in \R^{n_{layer}} $	& $ m $		& array of layer depths \\
$ (S_i)_{i=1}^{n_{layer}} $						& $ \in \R^{n_{layer}} $	& $ \permil $	& array of salinity \\
$ (T_i)_{i=1}^{n_{layer}} $							& $ \in \R^{n_{layer}} $	& $ {}^\circ C $	& array of temperature \\
$ ((y_{j,i})_{i=1}^{n_{layer}})_{j=1}^{n_{tracer}}$		& $ \in \R^{n_{layer} \times n_{tracer}} $	& $ m \, mol / m^3 $	& array of tracer concentrations \\
$ ((y_{bgc,j,i})_{i=1}^{n_{layer}})_{j=1}^{n_{tracer}} $	& $ \in \R^{n_{layer} \times n_{tracer}} $	& $ m \, mol / m^3 $	& computed source/sink array \\
$ n_{param} $									& $ \in \N $			& -			& number of parameters\\
$ (u_{i})_{i=1}^{n_{param}} $						& $ \in \R^{n_{param}} $	& -			& array of parameters
\end{tabular}
\end{center}


\subsection{Realization in C and FORTRAN}

In C real numbers are realized as \texttt{double} and natural numbers as \texttt{int}.
An array is represented by a pointer and its length is either the according \texttt{int}
argument (one-dimensional array) or the product of several corresponding \texttt{int}
arguments (multidimensional array).
Hence
\begin{verbatim}
void bgc(
    double t, double dt,
    double lat, double insol, double dayfrac, double ice, double wind,
    int ntracer, int nlayer,
    double *h, double *d, double *salt, double *temp,
    double *y, double *ybgc,
    int nparam, double *u
);
\end{verbatim}

\bigskip

In FORTRAN real numbers are realized as \texttt{real*8} and natural numbers as \texttt{integer}.
An arrays is denoted as a variable with one or more dimensions (ranks).
Its length is either its dimension (one-dimensional array) or the product of dimensions
(multidimensional array).
Hence
\begin{verbatim}
subroutine bgc(
    t, dt, lat, insol, dayfrac, ice, wind,
    ntracer, nlayer,
    h, d, salt, temp, y, ybgc,
    nparam, u
)
real*8  :: t, dt, lat, insol, dayfrac, ice, wind
integer :: ntracer, nlayer
real*8  :: h(nlayer), d(nlayer), salt(nlayer), temp(nlayer)
real*8  :: y(nlayer, ntracer), ybgc(nlayer, ntracer)
integer :: nparam
real*8  :: u(nparam)
\end{verbatim}


%
%
\section{PETSc Solvers}

The {\sc PETSc} library offers several variants of sophisticated nonlinear equation solvers.
Its two basic techniques are a globalized Newton method using a cubic interpolation as line search (LS) method and an adaption of the {\em Trust Region Method}\/ for nonlinear equations. Detailed descriptions on both
can be found in optimization textbooks, e.g. in \cite{DenSch96}.
Both approaches require solving a linear system of equations with the Jacobian of the nonlinear mapping $\phi$ introduced in Section \ref{sec:spinup}.


\subsection{Matrix-Free Linear System Solvers}

In many systems resulting from three-dimensional spatial discretizations of PDEs, the Jacobian is too big to even store it, specifically if it is dense. A remedy is to  use matrix-free
iterative Krylov subspace methods as {\em Conjugate Gradients} or {\em GMRES}, which are also supplied by {\sc PETSc}.

The Jacobian-vector product
needed in a single Newton (and also Trust Region) step can be approximated by finite differences, or alternatively be computed exactly using Automatic Differentiation.



\subsection{Inexact Computation and Compatibility of Stopping Criteria}

The linear system of equations is always solved with a prescribed accuracy, but tight tolerances for inexact residuals can lead to ''oversolving''.
Eisenstat and Walker described in \cite{EisWal1996} the fundamentals for
inexact (truncated) Newton methods, namely the way how to adjust inner (i.e. linear system solver) and outer (i.e. Newton or Trust Region iteration) stopping tolerances. This methodology is also already implemented in the  {\sc PETSc} library.



%
%
\section{First Exemplary Results}
\label{ExemplaryResults}

In the following we present different convergence results for the PO4-DOP and
PO4-DOP-PHY model. The computations were performed on a Linux cluster with
2.1 GHz AMD Barcelona CPUs. They show the behavior of a globalized
Newton-Krylov solver provided by the PETSc library in contrast to a usual spin-up, which can be interpreted as a fixed point iteration.

The ocean circulation is computed using twelve implicit and twelve explicit matrices, which are
linearly interpolated over 2880 time steps, corresponding to one year.
They represent monthly averaged diffusion and advection and were evaluated using
the MIT general circulation model (http://mitgcm.org/) with a latitudinal and longitudinal resolution of
$2.8125^\circ$ and 15 vertical layers. This resolution results in a vector length of 52749 per tracer.
More details on the configuration of ocean physics can be found in \cite{KhViCa05}.

The biogeochemical models consist of two euphotic layers and eight internal time steps per
one ocean step. They were implemented in FORTRAN and coupled to the ocean circulation using
the described interface.

The solution we are looking for is a fixed point for the mapping $\phi$ that integrates the
tracers forward in time for one year. The figures show the the Euclidean norm of the
difference between the concentrations before and after the integration
\begin{align*}
\| y - \phi(y) \|_2 \, ,
\end{align*}
where $y$ is the vector of all tracers.

The error of the spin-up, i.e. the fixed point iteration, is computed after every year of integration.
The convergence results of the nonlinear solver depict the error of the residual $y-\phi(y)$ after
every Newton step and the error of the Krylov solver in between. The former uses a line search
technique with cubic backtracking. The latter is a GMRES solver with a finite differences
approximation for the Jacobian vector product, a restart parameter of 200 and a modified
Gram-Schmidt orthogonalization. The tolerances for the Krylov subspace solver are chosen
depending on the residual norm after each Newton step, see \cite{EisWal1996}.
Except restart and orthogonalization standard PETSc settings are used.

\begin{figure}
\begin{center}
\includegraphics[width=12cm]{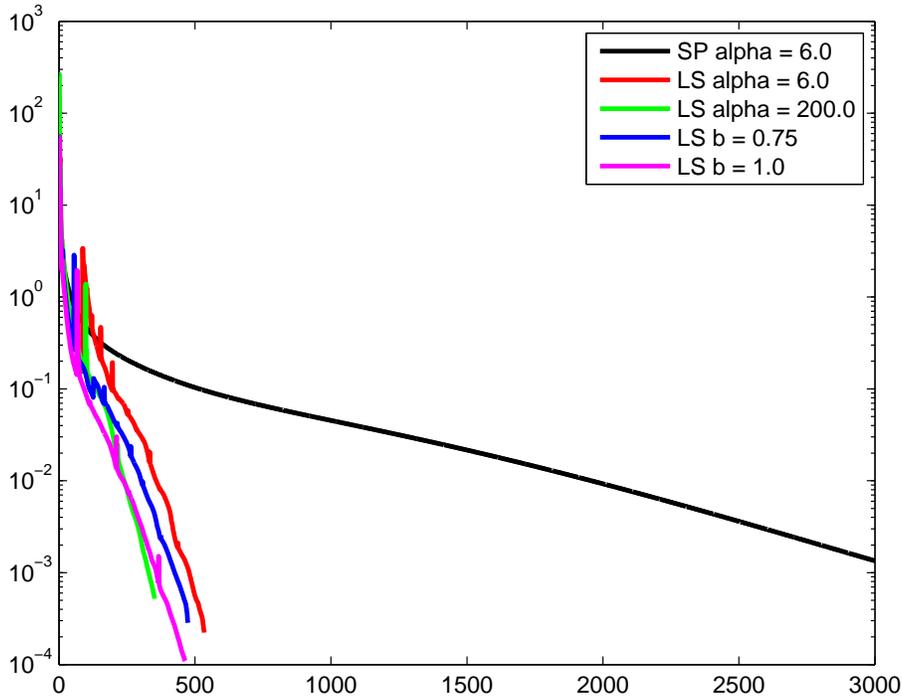}
\caption{\label{fig:PO4-DOP-conv}Comparison of the convergence towards
the solution of the PO4-DOP model by a spin-up (SP) and globalized Newton using a cubic line search (LS). The line search computations were performed
with different $\alpha$ and $b$ parameters.}
\end{center}
\end{figure}
\begin{figure}
\begin{center}
\includegraphics[width=12cm]{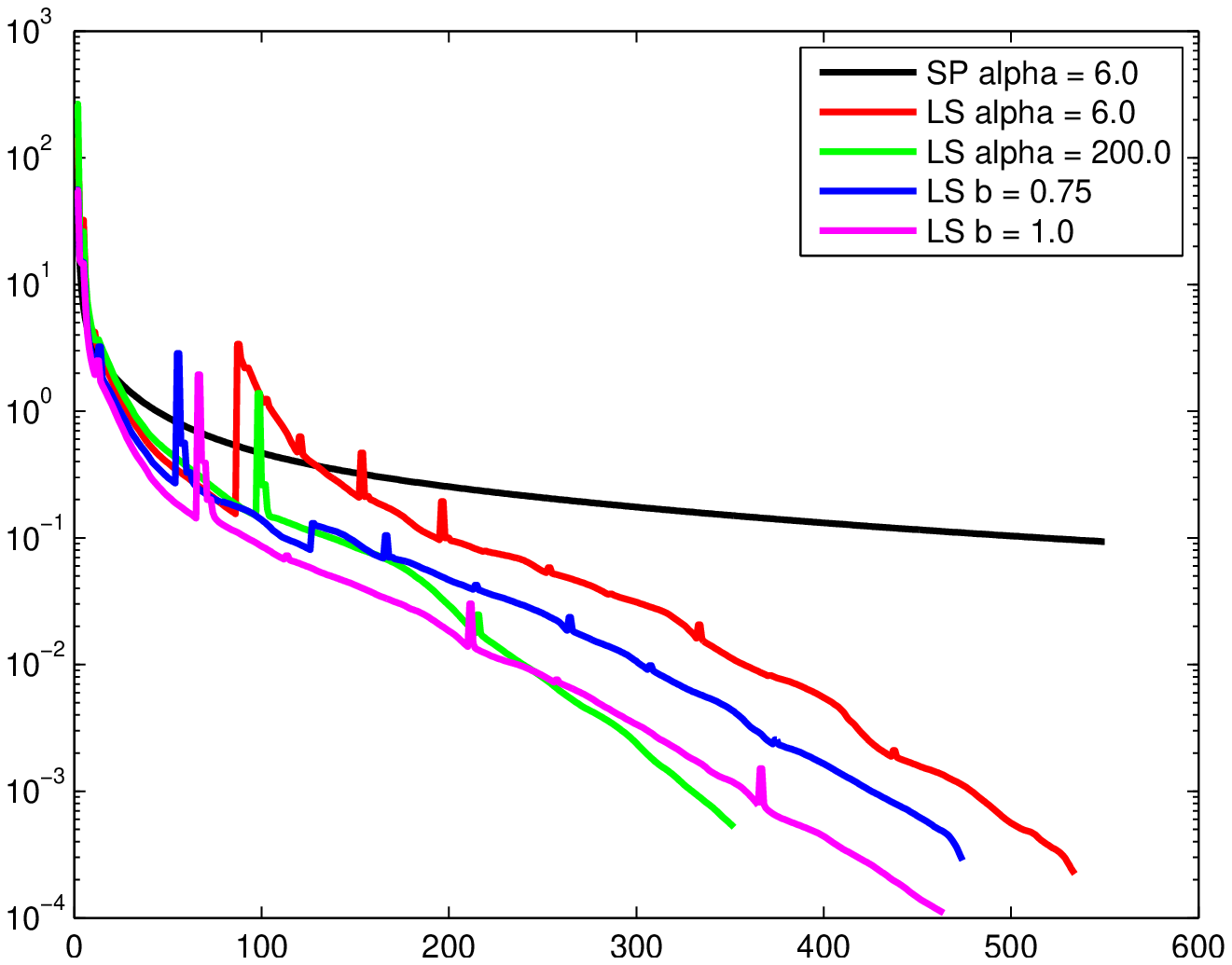}
\caption{Shows the first 550 model years of Figure \ref{fig:PO4-DOP-conv}.}
\end{center}
\end{figure}
%
%
\begin{figure}
\begin{center}
\includegraphics[width=12cm]{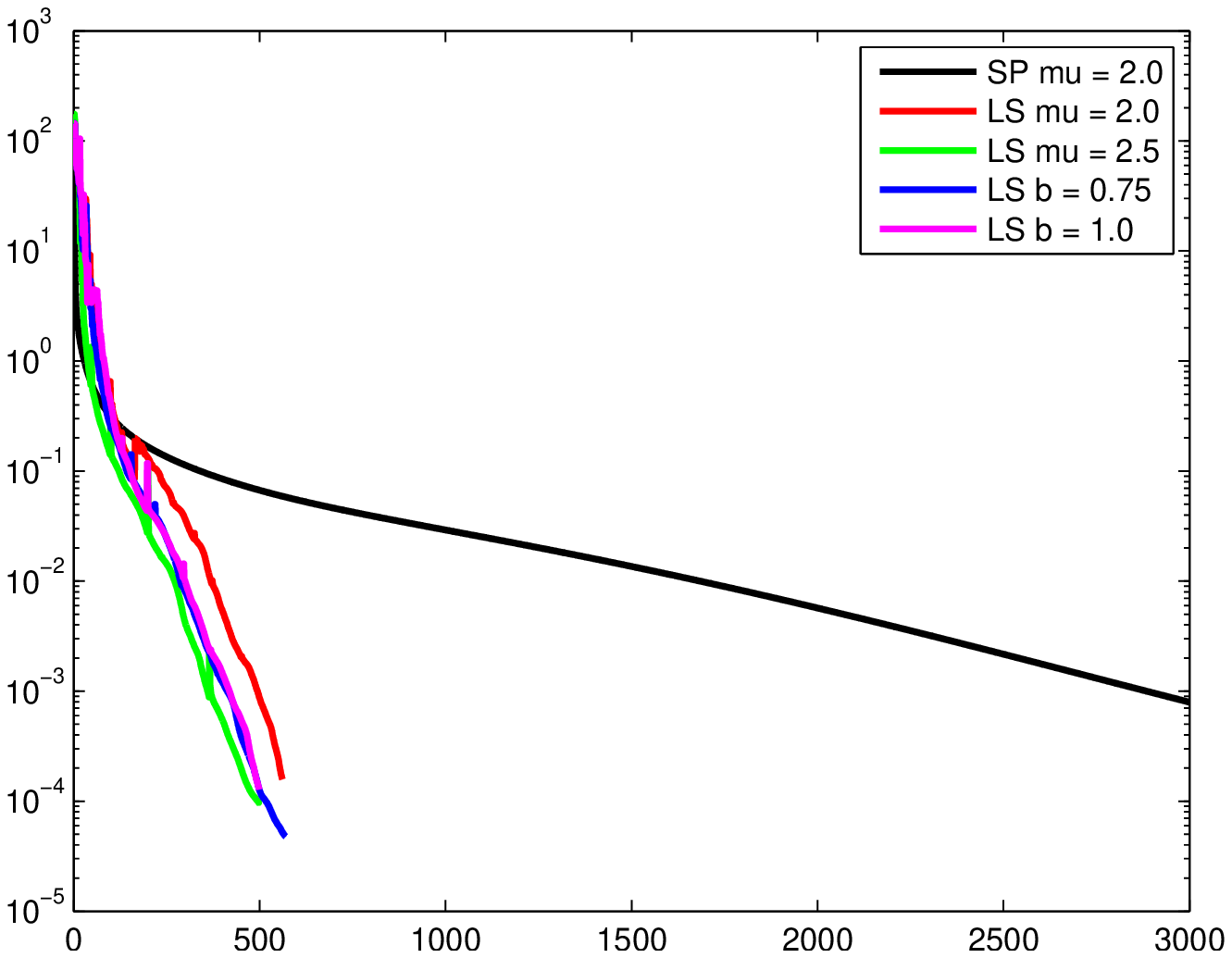}
\caption{\label{fig:PO4-DOP-PHY-conv}Comparison of the convergence towards
the solution of the PO4-DOP-PHY model by a spin-up (SP) and globalized Newton  using a cubic line search (LS). The line search computations were performed
with different $\alpha$ and $b$ parameters.}
\end{center}
\end{figure}
%
\begin{figure}
\begin{center}
\includegraphics[width=12cm]{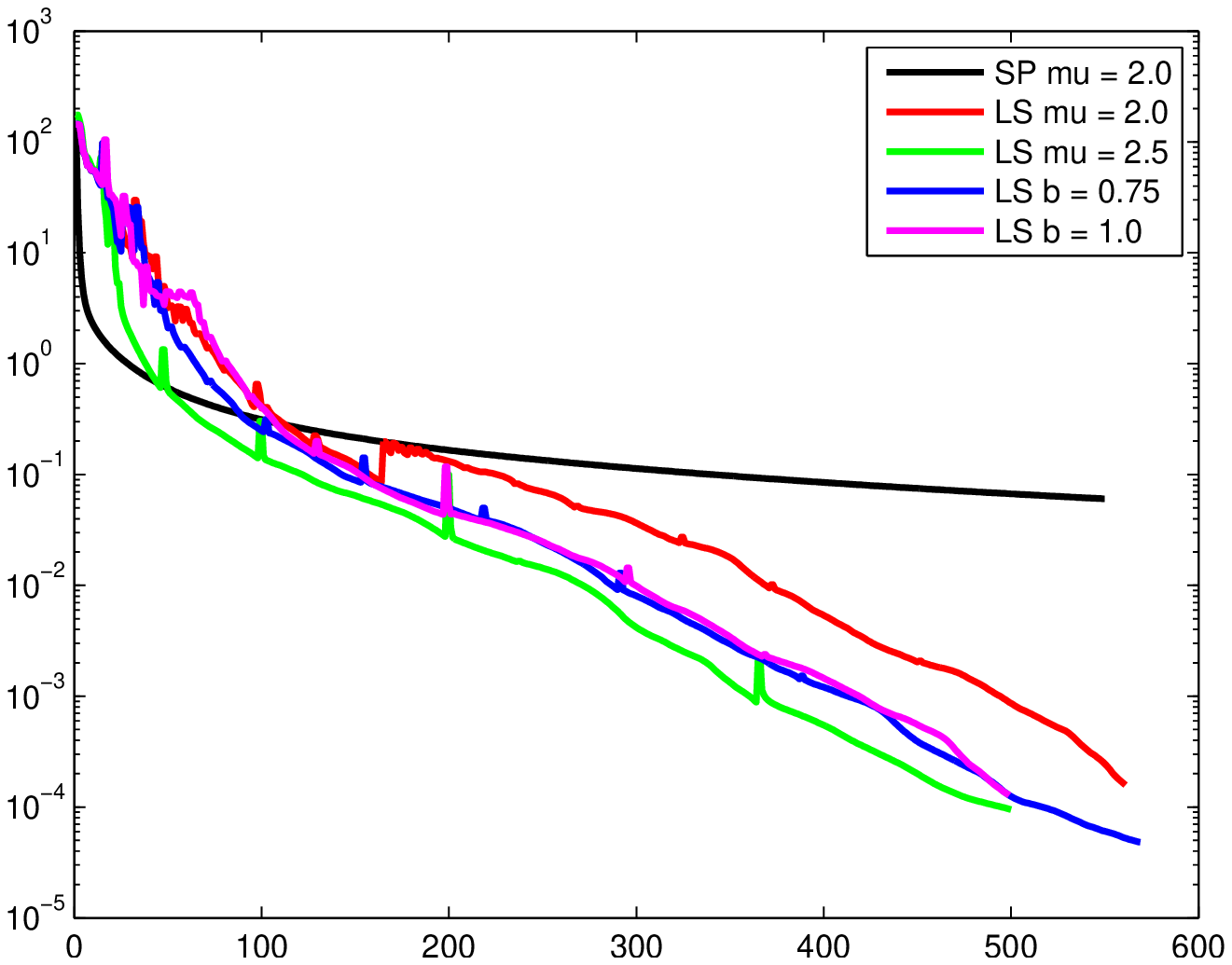}
\caption{Shows the first 550 model years of Figure \ref{fig:PO4-DOP-PHY-conv}.}
\end{center}
\end{figure}
%


%
%
\section{Flexibility and Portability}

Our entitlement is to provide a software that is as flexible and portable as possible.
Concerning the biogeochemical models we presented an interface which
is a first step towards this requirement.

Regarding the hardware there is demand for flexibility and portability too.
On the one hand it should be possible to make low resolution test runs on the home PC with
a multi core processor. On the other hand high resolution parameter optimization
clearly requires massively parallel hardware. Thus {\sc{Metos3D}} is based on PETSc.

First of all the library is open-source and freely available. It has been developed for
15 years now and is well maintained. Moreover there are over 200 applications,
extensively tested on different platforms, using PETSc.

The following figures show speedup test runs on different parallel hardware.
One is a Linux cluster (\texttt{rzcluster}) with 192 AMD Barcelona processors located at the University of Kiel.
The other is a massively parallel hardware ($\approx$ 20000 processors) located
in Hannover and Berlin. It is part of the North-German Supercomputing Alliance
(HLRN, http://www.hlrn.de/).

The tests were performed with a $2.8125^\circ$ and a $1.0^\circ$ resolution to
investigate the impact of the different load distributions on the speedup.

\begin{figure}
\begin{center}
\includegraphics[width=12cm]{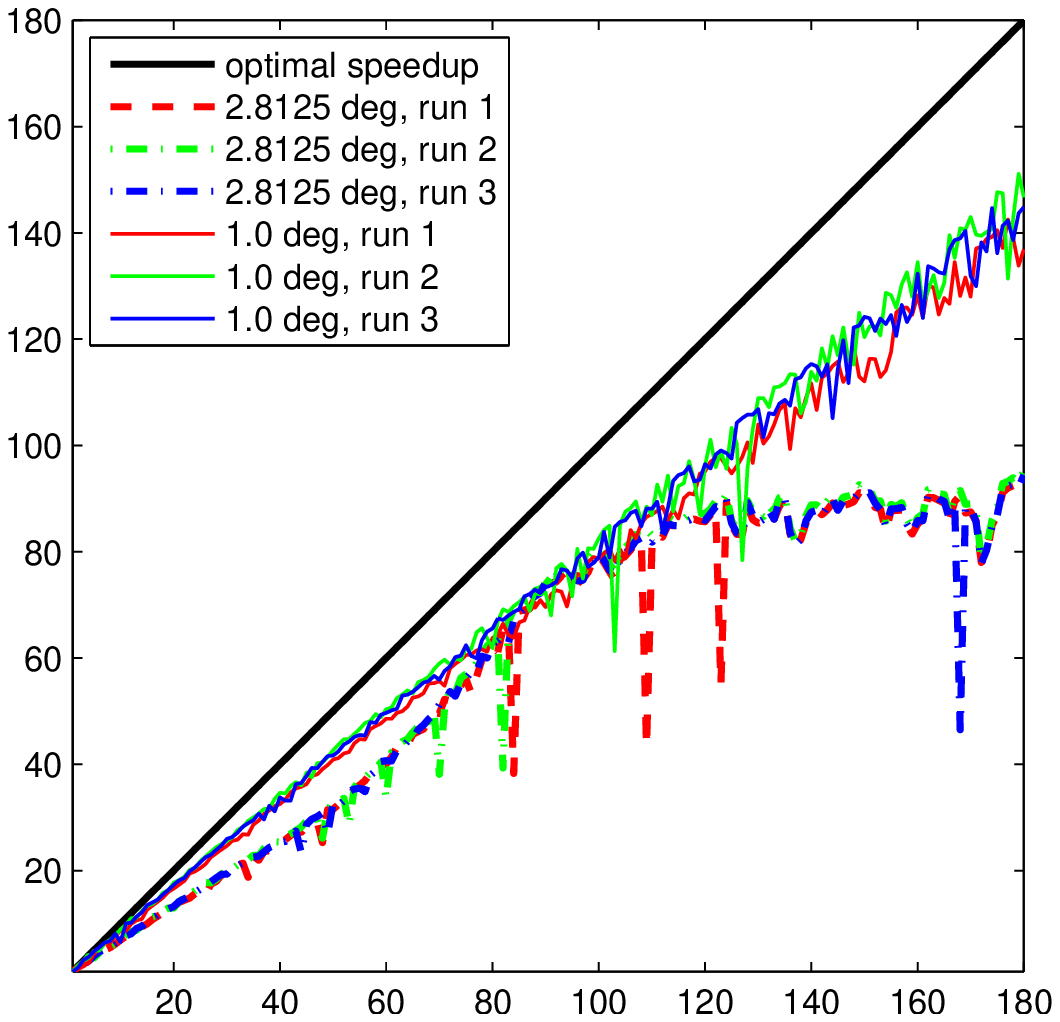}
\caption{Comparison of speedup for different resolutions on the \texttt{rzcluster}.}
\end{center}
\end{figure}
%
\begin{figure}
\begin{center}
\includegraphics[width=12cm]{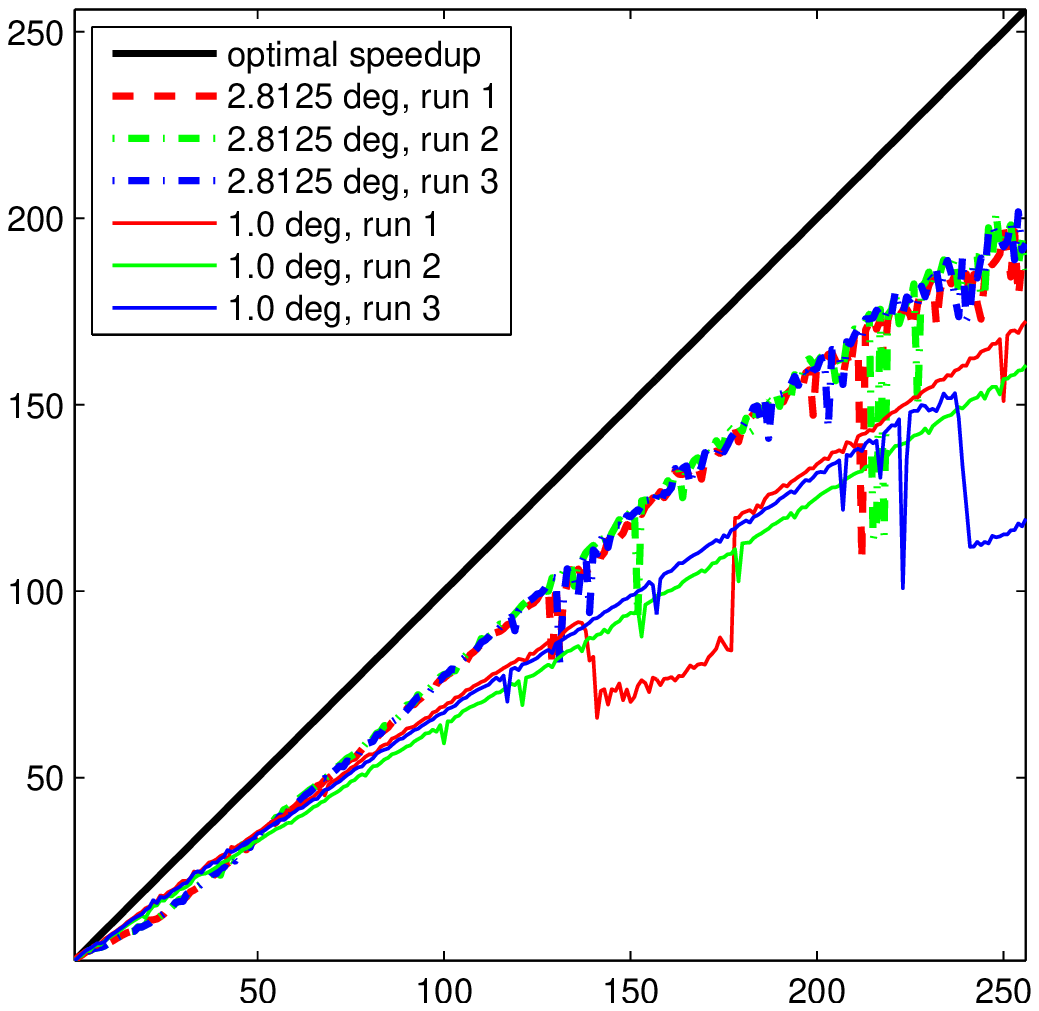}
\caption{Comparison of speedup for different resolutions on the HLRN.}
\end{center}
\end{figure}

%
%
\section{Next Steps}

As already pointed out in the title, this preprint is a first conceptual overview of the desired structure and features of {\sc Metos3D}.
We here additionallly  list some keywords that indicate the future direction of our work:
\begin{itemize}
\item 
Krylov subspace methods unfold their whole potential with a preconditioner, compare \cite{Kha08}, which is up to now not realized.

\item 
Moreover we can easily integrate  an interface for AD processed models/source code.
\end{itemize}


\newpage 

%
%
\bibliography{/Users/jpicau/Documents/ARBEIT/CODE/Literature/literature}
\bibliographystyle{plain}

\end{document}